# Conductance oscillations in graphene/nanoclusters hybrid material: towards large area single electron devices

*Florian Godel, Louis Donald Notemgnou Mouafo, Guillaume Froehlicher, Bernard Doudin, Stéphane Berciaud, Yves Henry, Jean-François Dayen\* and David Halley\**

Dr. F. Godel, L. D. N. Mouafo, G. Froehlicher, Prof. B. Doudin, Prof. S. Berciaud, Dr. Y. Henry, Dr. J.-F. Dayen\* and Dr. D. Halley\*
Institut de Physique et Chimie des Matériaux de Strasbourg, Université de Strasbourg, CNRS UMR 7504, 23 rue du Loess, BP 43, F-67034 Strasbourg Cedex 2, France
E-mail: dayen@unistra.fr, david.halley@unistra.fr

The fabrication of assemblies of particles having the same reproducible properties is of large interest for many domains in nanosciences, as for instance magnetic recording[1], catalysis[2], nanophotonics[3] or information technologies[4,5]. A driving motivation is to reproduce at the larger micrometric scale the dramatic physical effects arising at the single nanoparticle level. Targeted advantages include larger and robust signals, together with cheaper, simpler and scalable device fabrication process circumventing the technological bottleneck of contacting a single nano-object. This is a major issue for single-electron electronics, based on the discreteness of the electron charge, considered as a serious alternative to CMOS technology because of its very low power consumption and high-speed performance, with foreseen applications as sensors, memories, and multi-logic devices[4,6]. The key property in single electron device ('SED') is the discreteness of energy levels in metallic or semiconducting nanostructures, which results in well-defined Coulomb blockade ('CB') oscillations of the conductance when observed at the nanometer scale of a single nano-object, for example a nanocluster ('NC') or a molecule. However, despite those striking features, SED remains blocked at the stage of laboratory experiments. On the one hand, contacting and patterning a single NC with the required nanometric precision is a very challenging and expensive technological task. On the other hand, single electron features are smeared out when measured on large-scale devices, with potentially thousands of nano-objects contacted, having a distribution of size and electronics properties. There is therefore a need for new materials and/or device concepts allowing circumventing the problem of interfacing a single nano-object or of fabricating large numbers of perfectly identical NCs, while preserving the CB oscillations signatures at the core of SED concept.

Here we show that graphene ('Gr') is a promising platform to develop graphene-nanoclusters hybrid material ('Gr-NC hybrid'), while promoting a self-organized growth process with unique scalability and ease of processing, and demonstrating remarkable electronic properties for quantum electronics applications. Graphene was recently shown to be a potential candidate for the growth of homogenous and even monodisperse assemblies of metallic clusters[7–12] by exploiting surface modifications such as Moiré patterns, or by taking advantage of fast diffusion of metals over its surface[13,14]. We present a simple and scalable fabrication route of graphene-nanoclusters hybrid material, exploiting the self-organized growth over graphene of epitaxial flat Al nano-clusters covered by a thin Al-oxide tunnel barrier. We demonstrate here that graphene plays a dramatic role during the growth and oxidation of a very thin aluminum layer on its surface: when exposed to oxygen for a limited period of time, aluminum gets partially oxidized and metallic nanoclusters are left over the graphene inside the insulating oxide matrix. This specific growth mechanism confers to the Gr-NC hybrid material a structure ideally suited for single electron devices. Our data reveal that the conductance of Gr-NC hybrid behaves as if there were only a single contributing cluster or as if the assembly were almost perfectly monodisperse. This results in well-defined CB oscillations that are systematically observed on all samples, even in effective junction area as wide as 100 $\mu m^2$. This unique property is remarkably robust, as it was observed on two different types of graphene – exfoliated graphene and CVD graphene –, and allows making planar as well as vertical device architectures.

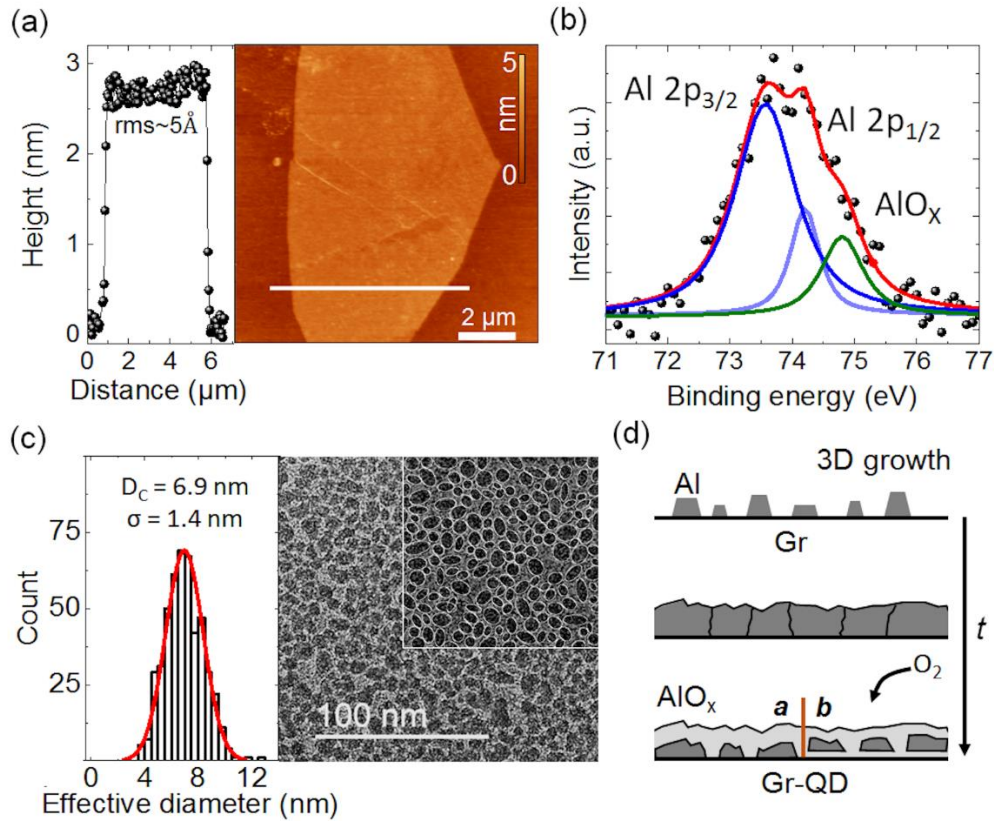

*Figure 1. (a) AFM image with deposited nominal Al thickness $t_{Al}$= 2.2 nm on top of an exfoliated 3 monolayer graphene film on Si0$_2$ with on its right the height profile corresponding to the white line with roughness below 5Å. (b) XPS spectrum performed ex situ on a sample made with $t_{Al}$= 1.6 nm on top of CVD graphene on Ni. (c) TEM picture for $t_{Al}$= 2.2 nm oxidized in an ambient atmosphere deposited on CVD graphene transferred on a carbon TEM grid. The corresponding histogram of the clusters' diameter D is fitted with a Gaussian curve centered at $D_C$=6.9 nm with a standard deviation σ of 1.4 nm (20%). (d) Sketch of the Al growth and oxidation in the case where (a) aluminum clusters are in contact with graphene or (b) surrounded by alumina.*

Samples fabrication rely on Gr-NC hybrids formed from commercial CVD graphene (4-7 layers) grown on polycrystalline nickel or from exfoliated graphene (1-3 layers) transferred onto amorphous $SiO_2$. In both cases, an aluminum layer with nominal thickness close to 2 nm was deposited by electron gun evaporation over graphene, oxidized in ambient atmosphere, and finally covered with a 40 nm thick cobalt layer capped with Au or Pd. Prior to aluminum deposition, Raman spectroscopy [15–19] was performed in order to ascertain the number of carbon monolayers and check their quality (**Figure S1**). Recent studies showed that, for nominal thickness below 1 nm, the growth of aluminum on graphene is three-dimensional[20,21] with flat clusters forming over the carbon sheet. In our case, we oxidize thicker Al layers, with nominal thickness between 1.5 and 3 nm: This thicker but still very thin layer allows us to provide a continuous layer without pinhole compatible with electronic device purposes, while keeping at the same time the 3D growth footprint through the grain boundaries. Atomic Force Microscopy (AFM) measurements indeed show that after aluminum oxidation, the graphene surface is rather smooth (**Figure 1(a)**), even on the micrometer scale, with a *rms* roughness close to 0.5 nm (resp. 0.3 nm) in the case of bilayer-graphene (resp. trilayer-graphene), and pinholes are not present. This confirms that aluminum is thick enough to form in both cases, after the likely 3D initial step[21], a continuous film on the underlying graphene[20,22]. In the case of CVD graphene, evidence for an epitaxial Al growth, on the scale of the Ni substrate grains, can moreover be observed in Reflexion High Energy Electrons Diffraction (RHEED) diagrams (**Figure S2**), which is consistent with the rather low lattice mismatch of 0.8 % between graphene and Al(111)[23].

Chemical, textural and structural analysis of the Gr-NC hybrid material were obtained by X-Ray Photoelectron Spectroscopy (XPS), Atomic Force Microscopy (AFM), Transmission Electron Microscopy (TEM) and micro-Raman measurements.

XPS was performed ex-situ on a 1.6 nm thick layer of aluminum deposited on CVD graphene on Ni and oxidized under ambient conditions (**Figure 1(b)**). This spectrum demonstrates that the oxidation of aluminum is only partial. Indeed, while a contribution of Al oxide is observed at 74.8 eV[24], clear contributions of metallic Al remain present at 73.6 eV ($2p_{3/2}$ photoemission peak) and 74.2 eV ($2p_{1/2}$ peak). A quantitative analysis of the data indicates (**Experimental Section**) that a relatively large fraction of the layer remains metallic, with less than 0.8 nm of Al transformed into oxide[25]. This is in agreement with results from the literature concerning alumina tunnel barrier formed by oxidation of Al films in air[26–28], where the oxide layer is usually found to be 0.5 to 0.8 nm thick.

Transmission Electron Microscopy (TEM) observations performed in plane-view (**Figure 1(c)**) reveals that an assembly of crystalline clusters is embedded in an amorphous matrix. The average lateral size of the clusters is 6.9 nm and the full-width-at-half-maximum of their size distribution, which is approximately Gaussian, is close to $2\sigma = 2.8$ nm. Notice that samples studied include multilayers and also single-layer graphene, excluding a NCs growth scenario via Al intercalation in-between graphene layers. Considering the smoothness of the oxidized film observed by AFM and the large remaining fraction of metallic Al deduced from XPS, we can propose a scenario for the clusters formation illustrated in **Figure 1(d)**: the Al clusters, which form at the beginning of growth, coalesce and form a continuous layer when the nominal thickness exceeds 1.5 nm; when exposed to oxygen, aluminum oxidizes from the top surface and the grain boundaries – where clusters coalesced –, leaving buried metallic Al clusters.

A comparative study of spatially-resolved micro-Raman measurements acquired on pristine graphene before and after the deposition/oxidation of aluminum is shown in **Figure 2(a)**. The resulting Raman maps are very similar, which demonstrates that Al deposition does not have detrimental effects on the graphene quality. First, no signature of additional doping or strain emerges, as illustrated by the nearly identical correlation plots of G- and 2D-mode frequencies[15–17] (**Figure 2(b)**). Second, no significant increase of the intensity of the defect-related D-mode feature relative to that of the G-mode feature is observed (**Figures 2(c)-(d)**), with a $I_D/I_G$ ratio remaining very low (below 0.1).

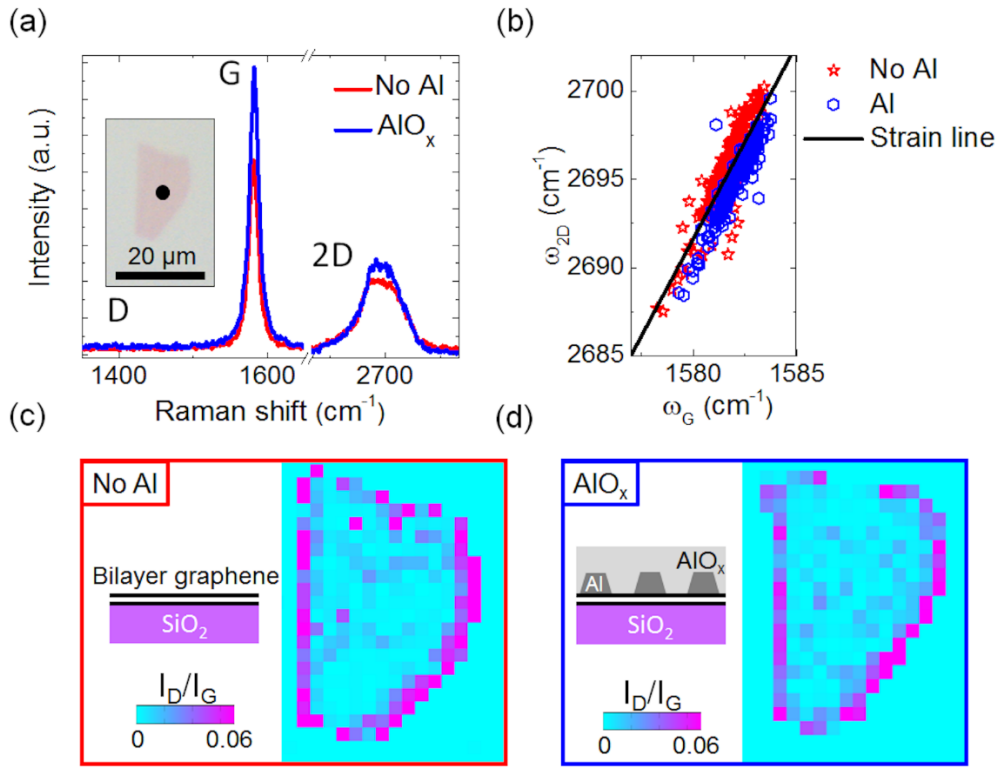

*Figure 2. (a) Raman spectra recorded with a 532 nm (2.33 eV) laser on bilayer graphene, before and after deposition of 2.2 nm of Al. Inset: Optical image of the corresponding bilayer graphene on $SiO_2$(285 nm)/Si substrate where the black dot represents the position where Raman spectra have been taken. (b) Correlation between the 2D-mode and G-mode frequencies before (red stars) and after (blue hexagons) aluminum deposition /oxidation. The black line represents the strain line with a typical slope of 2.2 (see text for details). (c, d) schematic view of the sample (left) and maps of the $I_D/I_G$ ratio (right) recorded on the same sample before (c) and after (d) Al deposition.*

It is important here to notice that the growth of metal thin film over graphene is subtle, as it was shown to be strongly dependent on the deposition technique[29–32], leading in some cases to dewetted films, or films with pinholes[30–32]. More works exploring the robustness of the hybrid growth process respect to experimental parameters such as pressure, surface roughness, and deposition technique will allow ascertaining if the hybrid growth process reported here could be extended to other experimental conditions. However, as it will be detailed now, all devices prepared according to this growth process, using either multilayer CVD graphene grown over Nickel substrate or exfoliated (mono or multilayers) graphene transferred over $SiO_2$, demonstrated similar conductance oscillations resulting from charge transport through Al nanoclusters, which is a strong indication that the hybrid growth process is fairly robust respect to these two graphene substrates.

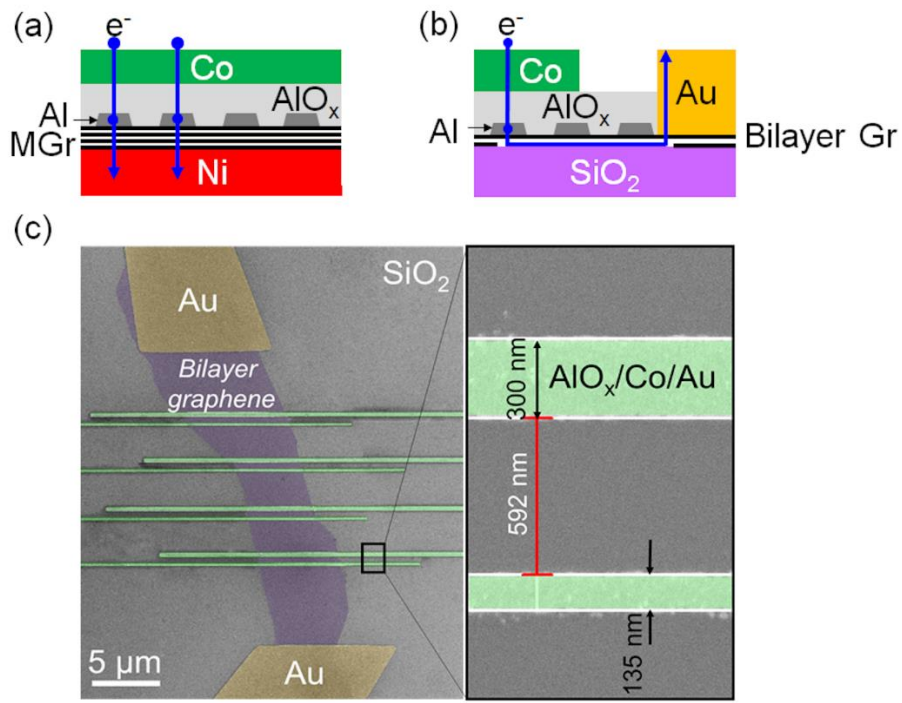

*Figure 3. (a) Schematics of the tunnel junction in the vertical configuration for CVD-grown substrates. (b) Planar device configuration on exfoliated graphene over insulating substrate. (c) SEM picture of a planar device (left) with detailed zoom on the electrodes (right).*

Study of the electronic and transport properties of the Gr-NC hybrid are performed on two complementary geometries, referred to as vertical and planar devices, made from either multilayer CVD graphene grown over Ni or graphene flakes exfoliated over $SiO_2$ respectively. The devices were fabricated by standard electron lithography processes (**Experimental section**): In the case of commercial CVD graphene on nickel, vertical tunnel junctions have been built by interconnecting the Gr-NC hybrid stack between the Ni layer used for CVD growth and a top Co electrode deposited over Al layer after its oxidation process (**Figure 3(a)**). In the case of exfoliated graphene on $SiO_2$, the current is also injected vertically from the cobalt top electrode into the Gr-NC hybrid stack, but charges are extracted laterally along the graphene film and collected through a side-patterned Au electrode (**Figures 3(b)-(c)**). Both geometries are complementary. The vertical geometry allows simple and large scale processing of devices, compatible with vertical integration scheme for low power and memory devices. The planar architecture has several added values and provides interesting perspective for future research opportunities. First this allows for designing several independent quantum boxes – while patterning several electrodes – that are interconnected through the graphene plane, offering the prospect of single-electron logic devices while driving single electron from one box to the others through the tunable graphene channel. Moreover, this should allow further development towards single-electron transistor based on this hybrid material while implementing a third terminal back gate control.

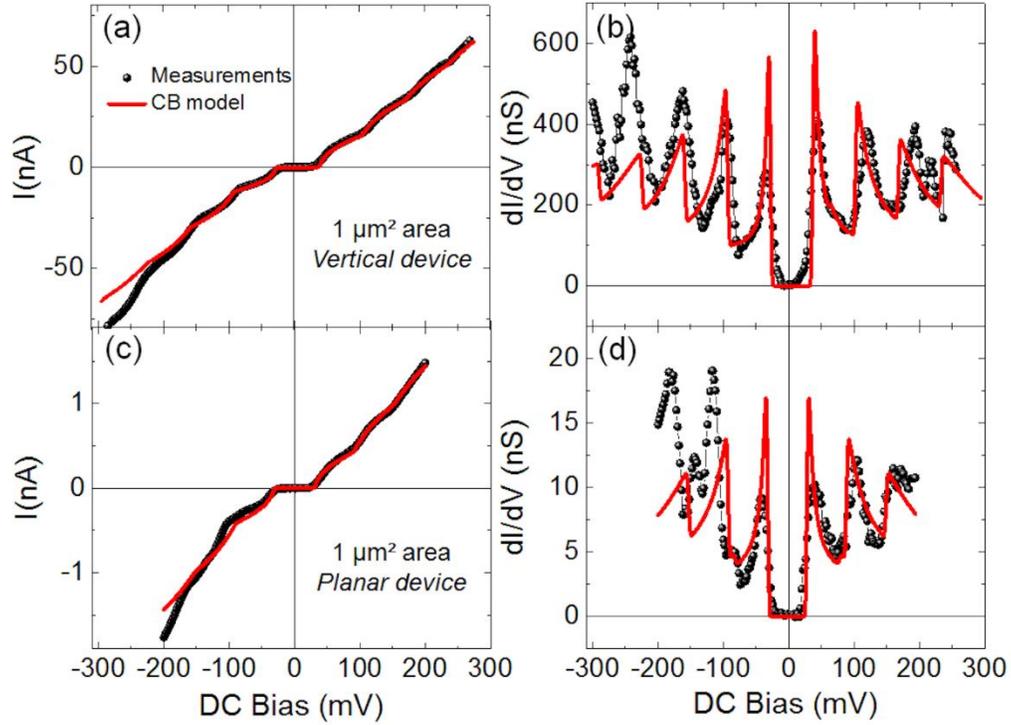

**Figure 4.** Experimental I(V) curves (black) measured at 1.5 K on a 1 µm² tunnel junctions with $t_{Al}$=2.2 nm (a) in a vertical device and (c) in a planar device configurations, with (b,d) corresponding dI/dV curves. Numerical simulations (red curves) assume a sequential Coulomb blockade model with $R_1$=6 MΩ, $C_1$=2.45 aF, $R_2$=1.5 MΩ, $C_2$=1 aF, $Q_0$=-0.1e for (a,b) and $R_1$=100 MΩ, $C_1$=2.6 aF, $R_2$=60 MΩ, $C_2$=1 aF, $Q_0$ = 0 for (c,d).

**Figure 4(a)** shows a typical current-voltage curve measured at low temperature (1.5 K) on a CVD graphene based vertical device with a cross-sectional area of 1 µm². This I(V) characteristic exhibits a well-defined Coulomb staircase and the corresponding differential conductance curve (**Figure 4(b)**) reveals narrow Coulomb blockade oscillations with a periodicity of 70 mV. A well-defined conduction gap is observed between the two lowest order conductance peaks, where the differential conductance drops to almost zero. Remarkably, the hybrid devices built in the planar geometry (with exfoliated graphene) show I(V) characteristics and Coulomb oscillations exhibiting the same features than those of the CVD-based devices

(**Figures 4(c)-(d)**). As they use identical growth and oxidation conditions for the Al layer and are designed with comparable junction areas, they provide further evidence of the robustness of the growth process. **Figures 5(a)-(b)** illustrate the fact that the Coulomb-blockade induced gap and conductance peaks survive well above 1.5 K and up to several tens of K. This is in agreement with the NC charging energy of 70-100 meV inferred from the periodicity of the Coulomb-blockade oscillations. Note that we also checked that the graphene channel was not degraded by the hybrid growth process, while performing transconductance and conductance measurement along the Gr channel contacted at its two extremities by Au electrodes (**Figure S3**).

Overall, *all* 20 studied devices made from 9 different samples (9 deposits of Al on distinct films of graphene with varying numbers of carbon monolayers and different substrates) demonstrate Coulomb blockade oscillations at low temperatures. These also include devices containing exfoliated single-layer graphene (**Figure S4**). The NCs involved in Coulomb blockade may therefore sit directly on top – and not below – of the graphene layer, which possibly forms the second (bottommost) tunnel barrier required for observing Coulomb-blockade. A thin oxide tunnel barrier at the interface with graphene may be formed, should Al oxidation occurs also from underneath. Previous electron transport studies[22,33,34] on tunnel junctions with thinner fully oxidized Al barriers on top of graphene did not report such Coulomb blockade resonant effects. This indicates that the mechanism responsible for the conductance oscillations in devices containing our Gr-NC hybrid material takes place in the Al clusters, or is induced in graphene by the presence of those clusters. The periodic conductance peaks that we observe look very similar to the Coulomb blockade oscillations observed historically by Ralph *et al.*[35] on single Al cluster in alumina and more recently on graphene by Coulomb blockade spectroscopy with a single metallic cluster attached to a scanning STM tip[36].

The fact that we observe similar CB features on the two kinds of graphene studied here is intriguing: multilayer CVD graphene, in the center of Ni grains, as well as exfoliated graphene have in common a very smooth surface, which maybe can explain that we observed in both cases the same behaviour in CB oscillations, *i.e.* similar mechanisms for the formation of aluminum clusters.

Numerical IV calculations based on the orthodox theory of single electron transport[37] provide more quantitative insight into the observed conductance oscillations. Alumina and graphene[38] are modelled as tunnel barriers for vertical charge transport via the metallic Al clusters. We note that this model could describe just as well a situation where the Al clusters would not sit directly on top of graphene but would be fully embedded in alumina. The simulations shown in **Figure 4** are performed assuming transport through a single Al cluster. The model therefore contains five adjustable parameters only (**Experimental section and Figure S5**): the resistances $R_i$ and capacitances $C_i$ of the two tunnel junctions (i = 1,2) and the background charge $Q_0$ induced by the NC electrostatic environment. As illustrated in **Figure 4**, this simple model reproduces fairly well the I(V) and dI/dV features, and, in particular, the position of the conduction peaks and the occurrence of a conductance gap. The capacitance values extracted from these simulations provide an estimate of the size of the clusters responsible for the Coulomb blockade oscillations. Assuming that the Al-oxide above the cluster is 0.8 nm thick[26] and that the cluster has the shape of a disc, we can estimate the contributing cluster diameter using a simple cylindrical capacitor model with $d = \sqrt{(4tC_1)/(\pi\varepsilon_0\varepsilon_{AlOx})}$ where $t$ is the AlO$_x$ thickness, $\varepsilon_0$ the vacuum permittivity and $\varepsilon_{AlOx}$ the dielectric constant of aluminum oxide taken as 9. We obtain a diameter equal to 5.6 nm ± 0.1 nm (**Figure 4(b)**) (respectively 5.7 nm for device of

**Figure 4(d)**), which is fully consistent with the clusters size distribution observed by TEM (**Figure 1(c)**). The data presented in **Figure 4** clearly points to the tunneling transport of single electron through only a single Al cluster. Indeed, if we consider that all clusters – about $10^4$ clusters in a 1μm² junction with the size distribution presented in **Figure 1(c)** and a Gaussian standard deviation of 1.4 nm ($\Delta d= 20\%$) – contribute to the transport, the CB oscillations sharpness dramatically vanishes in simulated dI/dV(V) curves (**Figure S6**) and only the zero-bias gap is maintained. We can observe a similar effect smearing out the CB oscillations even in the case of a smaller standard deviation with $\Delta d= 5\%$ (**Figure S6**).

The simulations of **Figure 4** also give some indications on the bottom tunnel barrier (referred as "tunnel barrier 2" in our model in **Experimental section** and **Figure S5**) between graphene and the nanocluster: the fitted $R_2$ values given in the caption are lower than $R_1$ but not negligible relatively to it, which could be consistent with the formation of a very thin alumina layer below the metallic cluster, at the graphene interface, playing the role of a tunnel barrier (**Figure 1(d)**, scenario *(b)*). This supposition is also supported by the observation of Coulomb blockade effects even in the case of a monolayer graphene (**Figure S4**).

A remarkable feature of the Coulomb staircases reported in **Figure 4** is the abruptness of the transitions between successive steps and the resulting sharpness of the conductance peaks. Such features can only be observed in the ideal situation where $R_1 > R_2$ and $C_1 > C_2$ (or $R_1 < R_2$ and $C_1 < C_2$). Such configuration is hard to achieve in usual nanostructures because the capacitance and resistance of a tunnel junction have opposite variation respect to the tunnel barrier thickness[4], which favors rather the opposite device configuration with $R_1 > R_2$ and $C_1 < C_2$.

In our Gr-NC hybrid material, the conditions $R_1 > R_2$ and $C_1 > C_2$ are fulfilled. This gives some important indication on the nature of the two tunnel barriers: in the hypothesis where a bottom alumina layer would alone play the role of the bottom tunnel barrier – thinner than the top one – we should have $R_1 > R_2$ but $C_1 < C_2$ contrary to what we observe. This indicates that probably, due to the mismatch of the wave vectors of the electrons at the Fermi level in aluminum and in graphene[33,39,40], the interface between graphene and the cluster might play the role of a layer in series with the alumina barrier. Such an effect likely promotes the ideal configuration where $R_1 > R_2$ and $C_1 > C_2$ simultaneously. This might represent an important advantage of Gr-NC hybrids for developing future single-electron devices.

To question the influence of the size of the tunnel junction on the abruptness of the I(V) steps, we have also studied devices containing tunnel junctions with surface area of 100 µm$^2$, hundred-times larger than in the case of **Figure 4**. For these devices, well-defined conductance gaps can still be observed (**Figures 5(a)-(b)**), and clear conductance oscillations are preserved (**Figure 5**). Interestingly, for devices of the same composition, we observe Coulomb oscillations with comparable periods and a limited spread in the value of the resistance-area products (for the four devices in **Figures 5(c)-(d)**, these are 3.5, 4.1, 4.2 and 4.9 MΩ.µm$^2$ at 300 K under 100 mV applied bias voltage). The conductance peaks are less pronounced than for 1 µm$^2$ devices (**Figure 4**) and, more importantly, they no longer have the asymmetric shape characteristic of transport through a single NC. In the largest devices, we expect several NCs to contribute in parallel to the observed current. Nevertheless, the periodic features in the conductance curves indicate that only selected NCs with very similar characteristics might participate in transport. Should this not be the case, Coulomb steps would be totally smeared out and good reproducibility of the Coulomb oscillations (**Figure 5(d)**) could not be achieved.

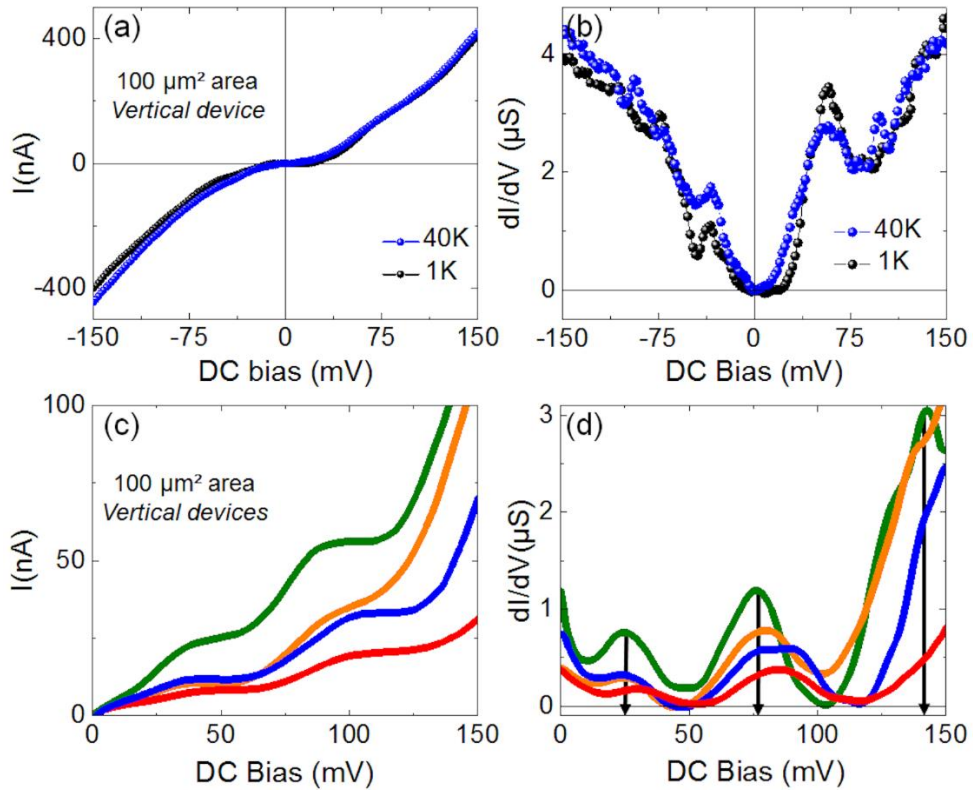

*Figure 5.* (a) I(V) and (b) dI/dV curves obtained at 1.6 K and 40K on a 100 μm² vertical device with $t_{Al}$=1.6 nm. (c) I(V) and (d) dI/dV curves obtained on four distinct perpendicular devices with 100 μm² junction area and the same nominal aluminum thickness ($t_{Al}$ =1.6 nm). Black arrows underline the comparable periods of the conductance oscillations for the four distinct devices. In (c) a linear current contribution attributed to direct tunneling between the electrodes has been subtracted [41].

Well-defined Coulomb staircases and reproducible single electron transport properties are very hard to obtain in devices as large as those studied here. So far, sharp Coulomb features could only be observed in cases where transport occurred through a single nano-object. This condition could only be met by either reducing the device area[41–46] to $10^2$ - $10^4$ nm², that is, up to 6 orders of magnitude smaller than our largest devices. For Al clusters embedded in Al-oxide, single NC operation in large devices could only be observed if hot spots are formed in relatively thick (> 20 nm) and rough insulating layers[26,28,41,44]. To our knowledge, CB oscillations have never been reported for large junctions based granular devices with thin and smooth tunnel barriers[27], such as used here[47]. Furthermore, the rare devices relying on hot spots do not exhibit reproducibility comparable to ours.

Our study revealing robust Coulomb blockade properties in devices based on the Gr-NC hybrid material raises a number of theoretical and experimental questions for future works. The first and most important one is: What property of the Gr-NC hybrid allows selecting NCs with very similar single electron transport properties among the assembly of metallic clusters? From the size distribution of the clusters, which is rather narrow and unimodal (**Figure 1(c)**), there is no evidence for the existence of a particular population that could dominate transport. Local tunnel conductance is however not solely determined by the size of the cluster involved. Interestingly, structural defects in graphene have been shown to constitute preferential sites for the nucleation of Al islands[21]. A possible explanation for the selection mechanism is that the height or the width of the bottom tunnel barrier is reduced for clusters lying above the graphene defects, thus promoting electron transmission at their locations. No link has however been established between preferential nucleation sites and cluster size.

Another question should be raised: could there be an electrostatic interaction between clusters, through a modulation of the electronic density in graphene[48]? In such a case, could it contribute to equalize the $Q_0$ values on the different clusters?

Another interesting question is whether similar hybrid materials could be formed using other 2D materials as templates. It could be the case since Coulomb blockade signatures have recently been observed in BN-based tunnel junctions with about 1 $\mu m^2$ cross-sectional area, capped with Cr/Au[49]. Unfortunately, no systematic structural analysis was performed and the phenomenon can therefore not be ascribed to the clustering of metal onto or into h-BN.

In conclusion, thin Al layer deposited on top of graphene and let to oxidize under ambient conditions forms a hybrid material of great interest, where flat nanometric Al clusters surrounded by Al-oxide sit over the graphene surface. The perpendicular injection of electric current through this Gr-NC hybrid material yields well defined periodic oscillations of the electric conductance, associated with Coulomb blockade thresholds, even in devices containing millions of clusters. These unique electronic properties are remarkably robust to the two types of graphene used (CVD or exfoliated), the number of layers involved (from 1 up to 7), or the substrate (conducting or dielectric). Our results suggest that the Gr-NC hybrid material has the property of selecting similar nanocrystals among a slightly disordered assembly so that the overall response of extremely large devices still exhibits sharp Coulomb blockade features.  Whatever the mechanisms at play, this result paves the way to future developments of single electron devices. Further systematic studies on the nominal aluminum thickness – *i.e.* on the clusters size – could optimize the CB resonant effects and hopefully make the CB oscillations observable at room temperature. We expect that this study will stimulate both experimental and theoretical efforts towards fine modeling and understanding of transport mechanisms in Gr-NC hybrid materials.

Different experimental developments could be followed: it would be for instance appealing to exploit the Moiré effect in 2D materials – graphene or boron nitride – which leads to a periodic modulation of the surface[50–52] and could support an organized self-assembling of the clusters or molecular species contributing to CB effects. Gr-nanoclusters hybrids are also promising for single electron spintronics: using ferromagnetic electrodes of Co or Ni as in our case, we could exploit the single charge effects combined with spin effects. This should make possible the observation of magneto-Coulomb effects[41] or spin accumulation in the metallic clusters[44]. Moreover, this could be combined with the unique properties of graphene in which we could inject and drive spins while benefiting from its unique long spin diffusion length[29,30,53,54] and spin filtering properties[33,39,55,56], and with implementing a back-gate terminal to tune the spin injection. Hence, Gr-NC hybrid offers promising prospects for the emergence of quantum devices based on graphene nano-heterostructures and graphene quantum dots[57,58].

Experimental Section

*Graphene samples*: We used two types of graphene. For vertical devices and XPS measurements we used 10 mm by 10 mm commercial substrates provided by Graphene Supermarket INC, covered by 4 to 7 layers of graphene grown on (111) textured nickel by chemical vapour deposition[55]. The average size of the Ni grains and of graphene domains varies from a few microns to almost 100 µm on a given substrate. It was already shown that graphene can efficiently protect nickel from oxidation[33,55], but this passivation effect probably depends on the interaction between graphene and the underlying metal in the case of Cu[59]. We therefore had to check in our case the absence or presence of Nickel oxide. XPS spectrum was thus performed on Al(1.6nm)/Gr/Ni sample after Al oxidation (**Figure S7**) in ambient atmosphere. The XPS spectrum confirms the Ni metallicity and its passivation by the graphene layers. No $NiO_x$ related peaks are indeed present[60,61].

For planar devices we used monolayer to three layers' thick graphene flakes (characterized by Raman and Atomic force microscopy) exfoliated mechanically with scotch tape on 285 nm silicon oxide/Si wafers.

*Devices fabrication*: Aluminum was deposited over graphene by electron gun evaporation in a high vacuum evaporator chamber or in an ultra-high vacuum Molecular Beam Epitaxy set up at a constant rate of 0,05 Å/s at room temperature and then oxidized in ambient atmosphere at room temperature. The devices exhibited identical performances and Coulomb peaks characteristics irrespective of the growth chamber and oxidation atmosphere, demonstrating the robustness and reproducibility of the growth mechanism. Vertical and lateral junctions were fabricated using electron beam lithography. For lateral devices, electrodes on exfoliated graphene are first patterned in spin coated PMMA resist and developed in 1:3 MIBK-IPA at 25°C. Typical electrode width and length are 300 to 500 nm and 3 to 5 µm respectively. The Al thin film (with nominal thickness close to 2 nm) is then deposited by e-beam evaporation into the electrodes pattern, and let oxidized at ambient atmosphere for at least one hour. After partial oxidation, a 40 nm Co layer, followed by a capping layer (4 nm of Pd or Au) is deposited by e-beam evaporation on top of the alumina barrier. Additional drain contacts, to drive charges out of the graphene channel, are then patterned by e-beam lithography followed by deposition (e-gun evaporation) of Ti(3 nm)/Au(47 nm) and lift-off. For vertical devices, the junctions were patterned in PMMA, developed in 1:3 MIBK-IPA, after which the resist was baked at 100°C to ensure its reticulation and its use as an electrical insulator between the graphene and top electrode. The Al growth, oxidation, and the Co top layer deposition, were then performed in the same way as for the lateral devices.

*XPS*: The spectrum was recorded using a monochromatic Al Kα source. The analysis was performed using XPSPEAK41 software. A Shirley type baseline was performed to remove the background. The pass energy during the measurement was 50 eV and the spot size was 1 mm². The XPS experiment is realized in 90° source/detector configuration. The aluminum oxide layer thickness can be determined using XPS measurements from the Al 2p metallic and AlOx peak area intensities. The AlO$_x$ thickness is obtained using this formula:

$$t_{AlOx} = \lambda_o \sin(\theta) \ln\left[\frac{N_m \lambda_m}{N_o \lambda_o} \frac{I_o}{I_m} + 1\right] \quad (1)$$

where $\lambda_{o(m)}$ is the inelastic mean free path of photoelectrons going through the aluminum oxide (metal), $N_{o(m)}$ the volume densities of Al atoms in the oxide (metal) and $\theta$ the angle between the XPS source and the detector. Typical value of $\lambda_o$ and $\lambda_m$ have been already determined as 28 Å and 26 Å respectively and the $N_m/N_o$ ratio is around 1.5[25]. The variability of $\lambda$ values has been estimated from Strohmeier[25] at around 15%. From XPS spectrum presented in **Figure 1(b)**, we extracted the Al metallic peak area intensity (Al 2p$_{3/2}$ and 2p$_{1/2}$) $I_m$, equal to 15000 and the AlO$_x$ one $I_o$ equal to 3500. Using the previous formula, we obtain a nominal thickness of aluminum oxide $t_{AlOx}$ of 7.9 Å ± 0.1 Å.

*Raman spectroscopy*: Raman spectroscopy is performed under ambient conditions. Raman spectra are recorded using a home-built setup with a 40× microscope objective and a 532 nm laser, which spot size is ~1 µm² and which power is maintained below 1 mW to avoid photothermal effects. The scattered signal is dispersed onto a charge-coupled device (CCD) array using a single-grating monochromator. The resulting spectral resolution is about 1 cm$^{-1}$. The sample holder is attached onto a piezo-stage, allowing spatially-resolved Raman studies, as shown in **Figure 2**[62].

*Transports measurements*: Low-temperature electrical measurements were carried out inside a He-flow cryostat of 1.5 K base temperature, using a low current source-meter K2634B for low signal measurements, with +/- 0.1 pA precision at 1.5 K.

*Orthodox method*: The Coulomb blockade model that we used is based on orthodox theory and master equation in the stationary regime[37]. We modeled the total current (using Matlab software) by choosing the key input parameters which are (C$_1$, R$_1$) and (C$_2$, R$_2$) couples, representing the two tunnel barriers (1 for Alumina, 2 for graphene) and Q$_0$ the environmental charge. Q$_0$ is taken between -0.5 eV and 0.5 eV corresponding to the Coulomb diamonds edges.

For each voltage step, we determined the number of electrons in the cluster by calculating the sum of tunnel transfer rates, allowing us to retrace the complete I(V) curve. Conductance curve dI/dV is obtained using 3 points derivative.

Acknowledgements

The authors thank Corinne Ulhaq-Bouillet and Raoul Arenal for TEM imaging, Marine Bouthillon for TEM image analysis, Thierry Dintzer, Laurent Simon and Samar Garreau for their help during XPS data acquisition and analysis, F. Federspiel, the members of the StNano platform for their technical assistance during nanofabrication and AFM, and Fabien Chevrier for technical assistance. They also acknowledge fruitful discussions with Pierre Seneor and financial support from the NIE Labex and the Agence Nationale de la Recherche (QuanDoGra (ANR-12-JS10-0001), Labex NIE 11-LABX-0058_NIE within the Investissement d'Avenir program ANR-10-IDEX-0002-02 and grant Nano G3N 2012.

Supporting Information

## 1. Raman spectroscopy on graphene before aluminum deposition

Prior to aluminum deposition, to ensure the quality of graphene and to determine the number of layers of our samples, we performed micro-Raman spectroscopy studies (see methods) on both exfoliated and CVD/Ni graphene samples. **Figure S1(a)** shows the typical Raman spectra of mono- and bi-layer [1] exfoliated graphene on $SiO_2$. For both, the G-mode frequency is $\omega_G \approx$ 1581 cm$^{-1}$ and its full width at half maximum is $\Gamma_G \approx 13$ cm$^{-1}$. These values correspond to quasi undoped samples with very small built-in strain[2,3]. For mono and bilayer samples, the $I_D/I_G$ ratio is below 0.01, a value which illustrates the very good crystal quality of these samples[1,4]. For graphene on Ni substrates, we recorded Raman spectra on each growth patch (**Figure S1(b)-(c)-(d)**). Knowing that the number of graphene layers is connected to the Raman 2D-mode lineshape[5] and that the Raman G-mode intensity in few-layer (<10) graphene samples is roughly proportional to the number of layers, we estimated that the number of graphene layers goes from 1 to ≈7, with some patches with 3 and 4-5 graphene layers. For both patches, the $I_D/I_G$ ratio is below 0.1, a value in line with other studies of graphene grown onto Ni films. Using the contrast as threshold between the different patches, we qualitatively reconstructed the optical image to give a spatial map of the number of graphene layers (**Figure S1(e)-(f)**).

## 2. RHEED characterization of aluminum

Reflexion High Energy Electrons Diffraction (RHEED) observations were carried out on a reference sample to avoid charging effects which could cause damages in the devices. When varying the positions of the incident beam on the surface or the azimuthal angle, we observe most of the times a diffuse and polycrystalline pattern, but surprisingly, for some positions and angles of the incident electron beam on the surface, the RHEED pattern shows the presence of relatively wide rods (**Figure S2**) together with a diffuse background. This suggests that, on the scale of a Ni crystalline grain, aluminum grows epitaxially on graphene[6,7]. In most of the cases, the Ni grains are too small and misoriented relatively to the incident beam, but in some cases, at least one graphene domain is large enough to contribute to the diffraction of the incident beam. This can be understood if we consider that the diameter of the electrons beam is in the order of tens of micrometers, close to the diameter of the largest graphene domains.

**3. Lateral conductance properties of the graphene channel.**

We have performed lateral transport measurement between Au electrodes patterned at the two extremities of the graphene flake of Gr-nanoclusters hybrids, which are in direct Ohmic contact with graphene (without the additional hybrid Al cluster/Al2O3 stack in between). As expected, graphene channels show usual transconductance (**Figure S3(a)**) and conductance (**Figure S3(b)**) behaviours. These results confirm that the graphene channel is not degraded by the hybrid stack growth, and that the conductance oscillations observed through the Gr-nanoclusters hybrids are due to the Al clusters present into the hybrid stack.

**4. Coulomb blockade features on graphene monolayer device**

Using planar device with a nominal aluminum thickness of 2.2 nm made on single-layer of graphene, we performed electrical characterization at low temperature (1.5 K). Well defined staircases are observed in both voltage polarizations (**Figure S4**). The dI/dV(V) curve in inset confirms the Coulomb blockade oscillations.

**5. Electrical circuit modeling of the device**

The numerical simulations of the electronic transport curves reported in the manuscript are obtained within an orthodox Coulomb blockade model (**Experimental section**), supposing an equivalent electrical circuit as shown in **Figure S5(b).**

**6. Diameter size influence onto CB features**

**Figure S6** shows the dI/dV(V) curves recorded at 1.5K on 1µm² vertical Co/AlOx/Al/Ni tunnel junction as presented in **Figure 4(b)**. Using orthodox theory, we perform simulation of CB transport considering different assembly of contributing clusters. The simulation supposing one single cluster, which fairly reproduces our data, is shown as red curve. We first considered clusters with a mean diameter of 5.6 nm: the simulation (blue curve) shows that CB peak's sharpness vanishes and does not reproduce experimental data. This effect is drastically enhanced if now all clusters are considered in the simulation; the CB oscillations are smeared out and the CB effect is no more reproduced, except the zero bias gap. These simulations point a tunnelling transport due to one selected size of clusters in such 1µm² junction.

## 7. Nickel passivation by graphene

**Figure S7** shows the XPS spectrum recorded using Al Kα source on graphene/Ni substrates after 1.6 nm Al deposition and oxidization in ambient atmosphere. The spectrum shows the Ni $2p_{3/2}$ peak originating from the Ni film under graphene. The typical signatures of metallic Ni are clearly visible with a $2p_{3/2}$ peak located at 852.6 e.V and its two related loss plasmons (Sat1 at 856.3 e.V (+3.7 e.V) and at Sat2 at 858.6 e.V (+6 e.V)[8]). No significant Ni-oxide signatures such as NiO peak expected at 854 e.V or the Ni $2p_{3/2}$ to Sat2 difference peak's energy is reduced to 5.8 e.V are not detected[9], thus confirming the passivation of the Ni layer by the graphene[10]. Another peak is present at 853 e.V which is directly connected to the Ni-C bond, coming from the carbon diluted atoms in the Ni film[11].

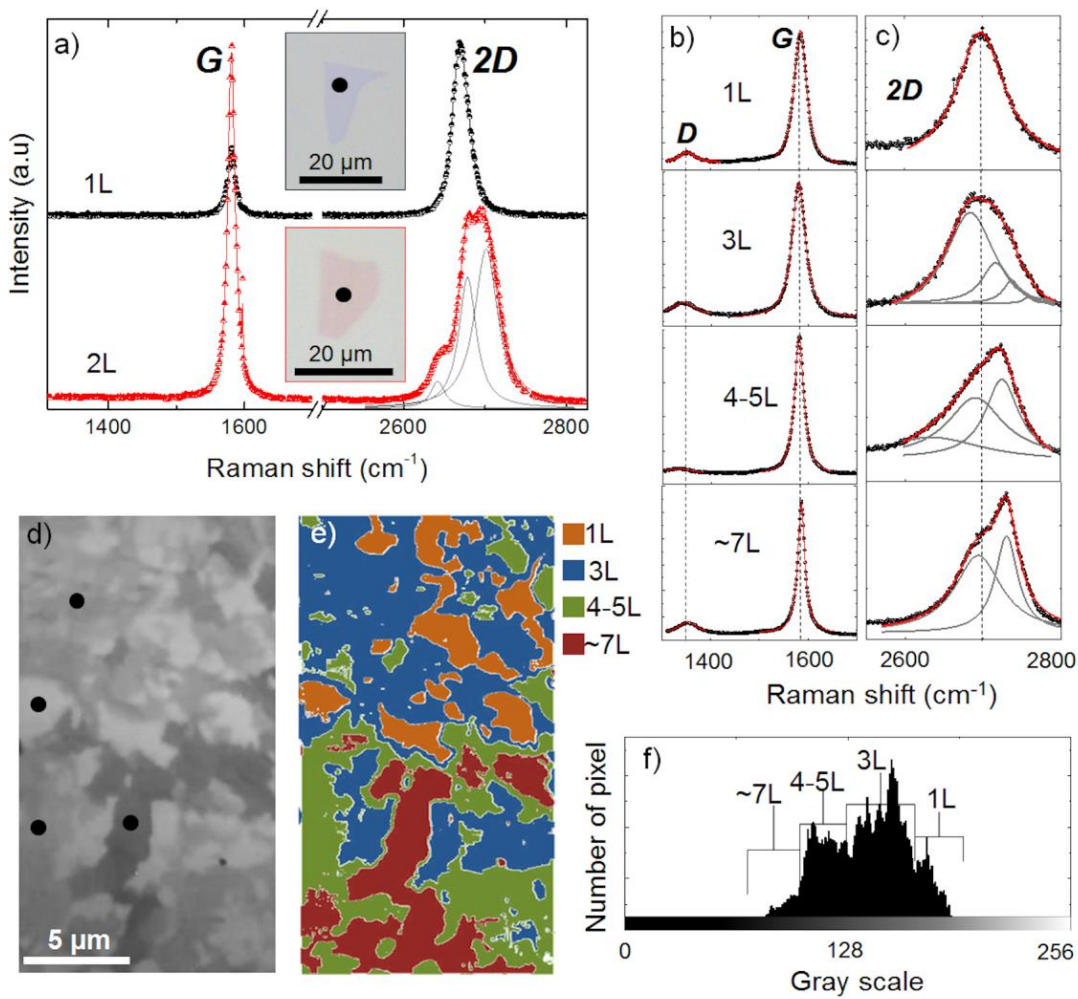

**Figure S1.** (a) Raman spectra recorded on mono- and bi-layer exfoliated graphene onto $SiO_2$(285nm)/Si substrate before aluminum deposition. Inset, the black (respectively red) frame is the optical image of the mono- (resp. bi-) layer of graphene. (b) D, G and (c) 2D peaks of different Raman spectra recorded on CVD graphene/Ni, corresponding to different patch with different numbers of graphene layers. The spectra in (a-c) are recorded at the black dots location on the optical image (d). (d) Optical image of CVD graphene/Ni in gray scale where patches of different numbers of graphene layers are clearly visible. (e) Reconstructed (d) image qualitatively indexing, for each patch, the number of graphene layers by a color (1L-monolayer, 3L-trilayer, 4-5L, 4-5 layers, 7L- 7 layers). (f) Histogram of pixel of the (d) image used for thresholding of the reconstructed (e) image.

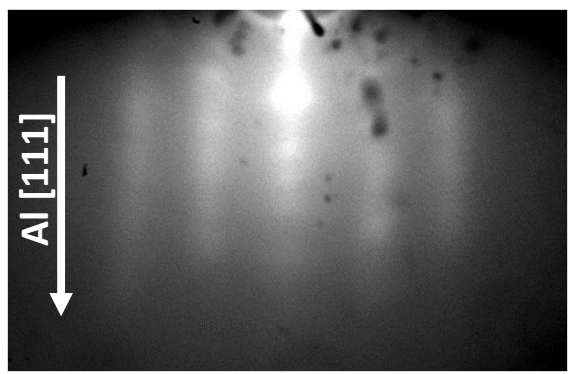

**Figure S2.** *In situ* RHEED diagram on 2.2 nm Al deposited on CVD graphene/Nickel before oxidation in ambient atmosphere.

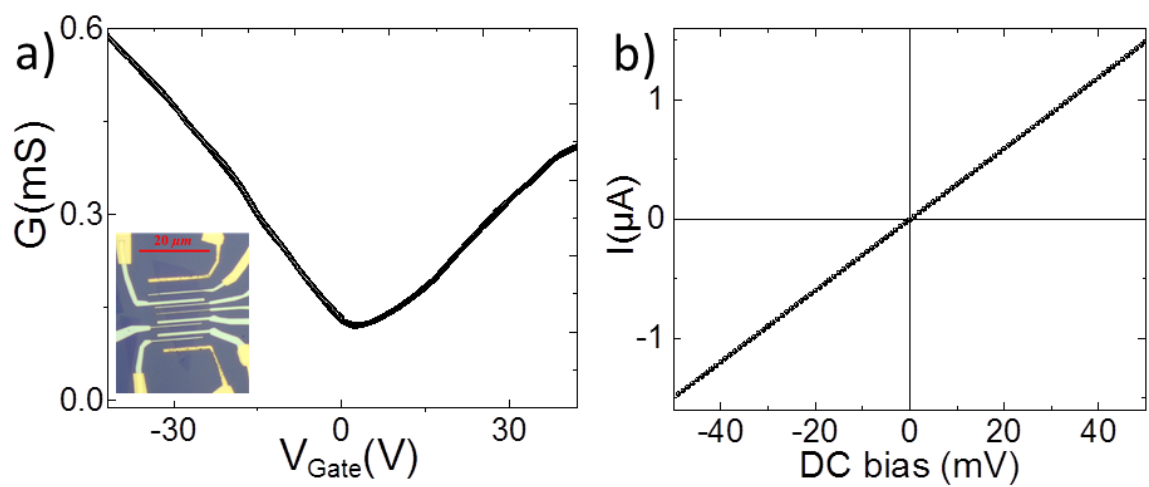

**Figure S3.** a) Typical transconductance curve of the graphene channel of a Gr-nanoclusters hybrid, measured between the Au electrodes at the two opposite sides of the channel. Inset:

optical image of the device, the yellowish electrodes are Au electrodes. b) Typical two terminals conductance curve of the graphene flake (of a Gr-nanocluster hybrid) measured between two Au electrodes.

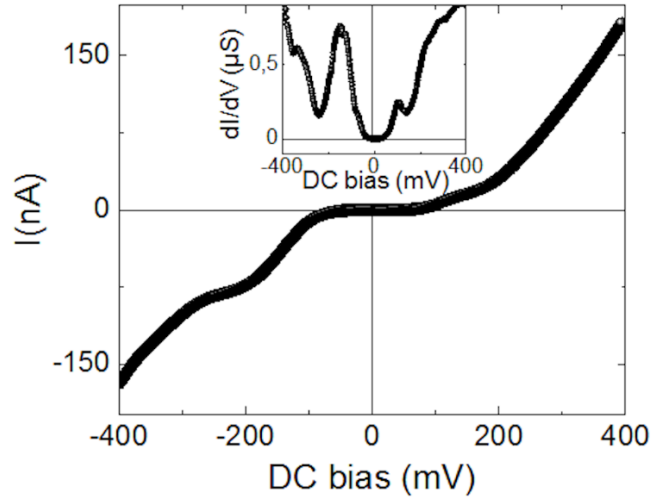

**Figure S4.** I(V) curve measured at 1.5 K on 1 µm$^2$ tunnel junction with $t_{Al}$=2.2 nm on top of a single layer exfoliated graphene in planar configuration. Inset, corresponding dI/dV curves with a low voltage gap and conductance oscillations.

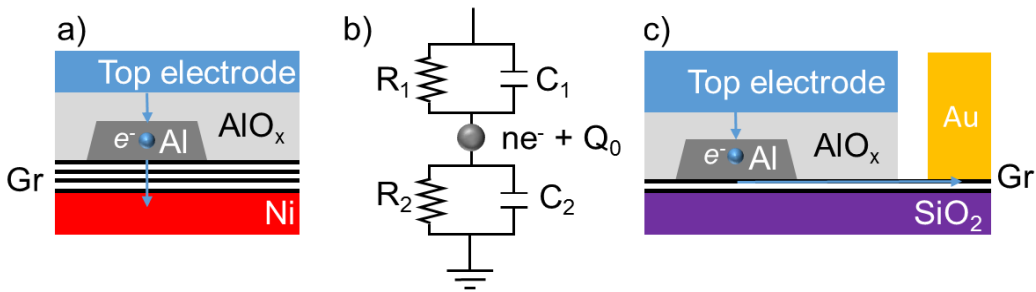

**Figure S5.** a) Schematic transverse view of the vertical devices. b) Equivalent circuit of the coulomb blockade system we used for the Orthodox theory calculation. ($R_1$, $C_1$) couple modelled the first barrier schemed by the dashed part of the alumina barrier. The other one, ($R_2$, $C_2$) modelled the second barrier embodies by the graphene layers. c) Schematic transverse view of planar devices.

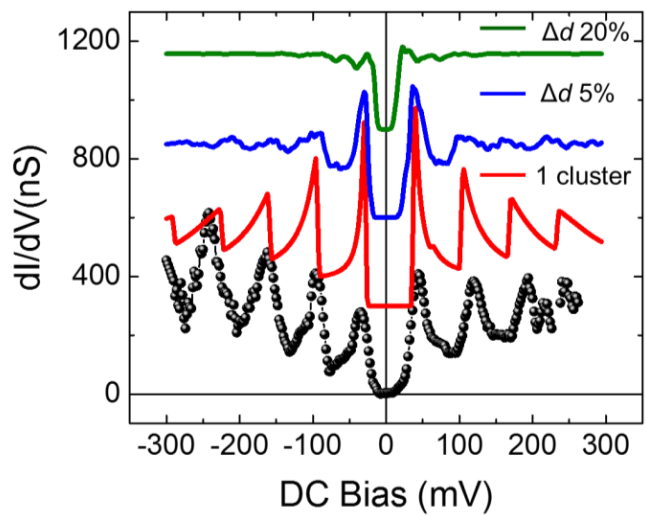

**Figure S6.** dI/dV(V) recorded at 1.5K on 1μm² vertical Co/AlOx/Al(2.2nm)/Gr/Ni junction (black dots) compared with different Coulomb blockade simulations: supposing one contributing cluster of 5.6 nm diameter (red curve), a distribution of contributing clusters which diameters are taken from a Gaussian distribution centered on 5.6 nm with a standard deviation Δd of 5% (2.8Å) (blue curve) and with a standard deviation of 20% (14Å) (green curve). Simulated curves are shifted for the sake of clarity.

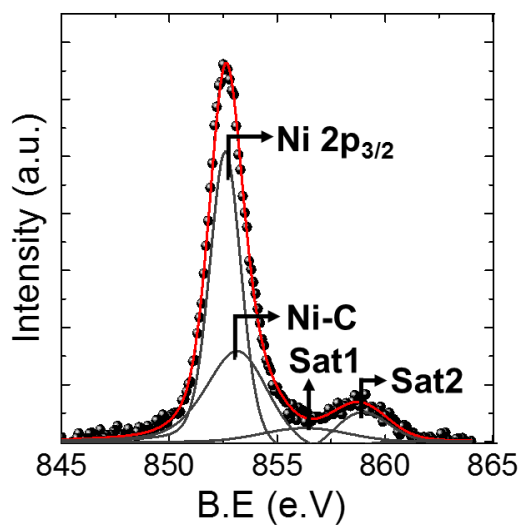

**Figure S7.** XPS spectrum recorded on Gr/Ni after 1.6 nm Al deposition and oxidization in ambient atmosphere. Spectrum is fitted using XPSPEAK41. A Shirley type baseline has been performed and removed from the recorded data presented here. The XPS spectrum fitted shape (red curve) comes directly from the contribution of each fitting by Gaussian-Lorentzian curve

type of the Ni $2p_{3/2}$ (852.6 e.V), Ni-C (853 e.V) peaks and Ni $2p_{3/2}$ first (Sat1 at 856.3 e.V) and second (Sat2 at 858.6 e.V) satellites.